\documentclass{iau} 
\usepackage{graphicx}
\usepackage{subcaption} 
\usepackage{float}

\title[JD 11.~~Short GRB Pulses and HMXRB Evolution] 
{How Pulses in Short Gamma-Ray Bursts \\ Constrain HMXRB Evolution}

\author[Jon Hakkila \& Robert D. Preece]   
{Jon Hakkila$^1$
 \and Robert D. Preece$^2$}

\affiliation{$^1$Dept. of Physics and Astronomy, College of Charleston, \\ 66 George St.,
Charleston, SC, USA \\ email: {\tt hakkilaj@cofc.edu} \\[\affilskip]
$^2$Dept. of Space Science, University of Alabama in Huntsville, \\ Huntsville,
AL, USA \\email: {\tt rob.preece@nasa.gov}}

\pubyear{2018}
\volume{346}  
\jname{High-mass X-ray binaries: illuminating the passage from massive binaries to merging compact objects}
\editors{A.C. Editor, B.D. Editor \& C.E. Editor, eds.}
\begin{document}

\maketitle

\begin{abstract}
We demonstrate how pulse structures in Short gamma-ray bursts (SGRBs), 
coupled with observations of GRB/GW 170817A, constrain the geometries of dying
HMXRB systems composed of merging neutron stars.
\keywords{X-rays: binaries, gamma rays: bursts, methods: data analysis, methods: statistical}
\end{abstract}

\firstsection 
\section{Introduction}

Binary neutron stars represent an HMXRB evolutionary end resulting in the creation of 
short gamma-ray bursts (SGRBs). These luminous flashes of $\gamma-$radiation occur
after neutron stars merge following the decay of their orbits. The most powerful evidence linking 
neutron star mergers to SGRBs has been the LIGO gravitational wave `chirp' of GW 170817
(\cite[Abbot et al. (2017)]{Abbot17})
in which two compact objects having masses between $1.17 M_\odot$ and $1.60 M_\odot$ 
and a combined mass of $2.74 M_\odot$ merged. SGRB 170817A was observed 1.7 seconds later by
the GBM experiment on Fermi. This SGRB had a duration of roughly 2 seconds 
(\cite[Goldstein et al. (2017)]{Goldstein17})
and a Lorentz factor of $\Gamma > 10$ (e.g., 
\cite[Zou et al. (2018)]{Zou18}).

The dominant method of emission in GRBs is via $\gamma-$ray pulses. 
GRB pulse light curves are not simple smoothly-varying `bumps.'
Instead they generally exhibit structure that cannot be
explained by stochastic background variations. Furthermore, this structure
often has a wavelike shape that gives a GRB pulse a triple-peaked
rather than a single-peaked appearance. Typical GRB pulses evolve from hard 
to soft but re-harden as the intensity re-brightens; this behavior is true for
both SGRBs and LGRBs (long GRBs).

\section{GRB Pulse Structure}

In order to characterize GRB pulse structure, \cite[Hakkila et al. (2018a)]{Hakkila18a} 
classified GRB pulses based on their complexity as determined by
a simple monotonic `bump' overlaid by an identifiable `wavy' structure; this simple approach is often effective. 
Pulses were classified as {\em simple} when they could be fitted by a monotonic pulse alone
(using the \cite[Norris et al. (2005)]{Norris05} pulse shape), 
{\em blended} when they could be fitted by a monotonic pulse with significant wavy residual structure
(characterized by the \cite[Hakkila \& Preece (2014)]{Hakkila14} residual function), 
{\em structured} when fits were improved but not completely adequate, 
and {\em complex} when pulse light curves were too structured for a good combined fit.

\begin{figure}[H]
\begin{center}
  \begin{subfigure}[b]{0.4\textwidth}
    \includegraphics[width=\textwidth]{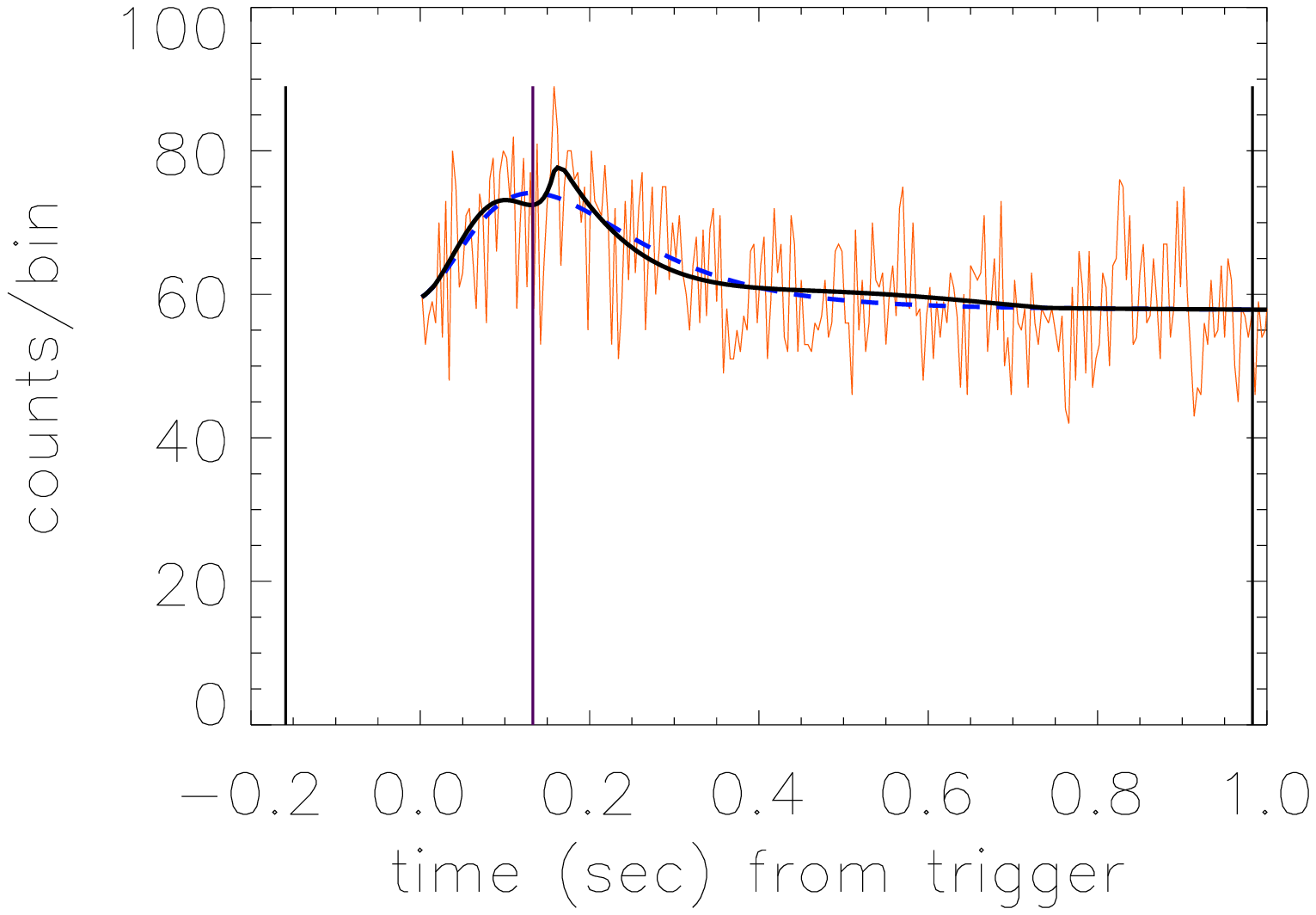}
    \vspace*{-0.5 cm}
  \end{subfigure}
  \begin{subfigure}[b]{0.4\textwidth}
    \includegraphics[width=\textwidth]{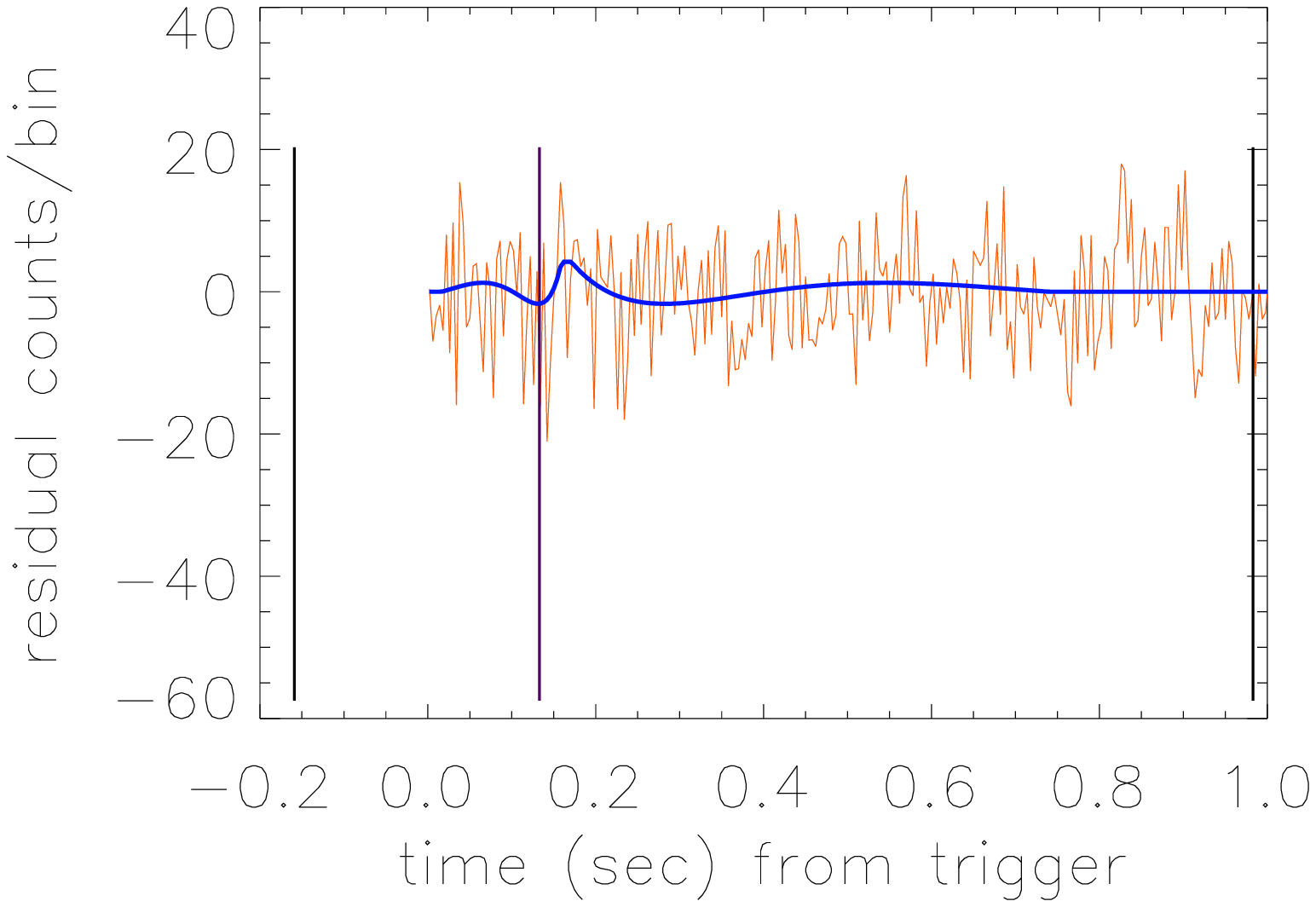}
    \vspace*{-0.5 cm}
  \end{subfigure}
  \end{center}
 \vspace*{-0.1 cm}
  \caption{Simple SGRB pulse BATSE 0373. Pulse fit (left) and residual fit (right).}
  \label{fig1}
 
\vspace*{-0.4 cm}
\begin{center}
  \begin{subfigure}[b]{0.4\textwidth}
    \includegraphics[width=\textwidth]{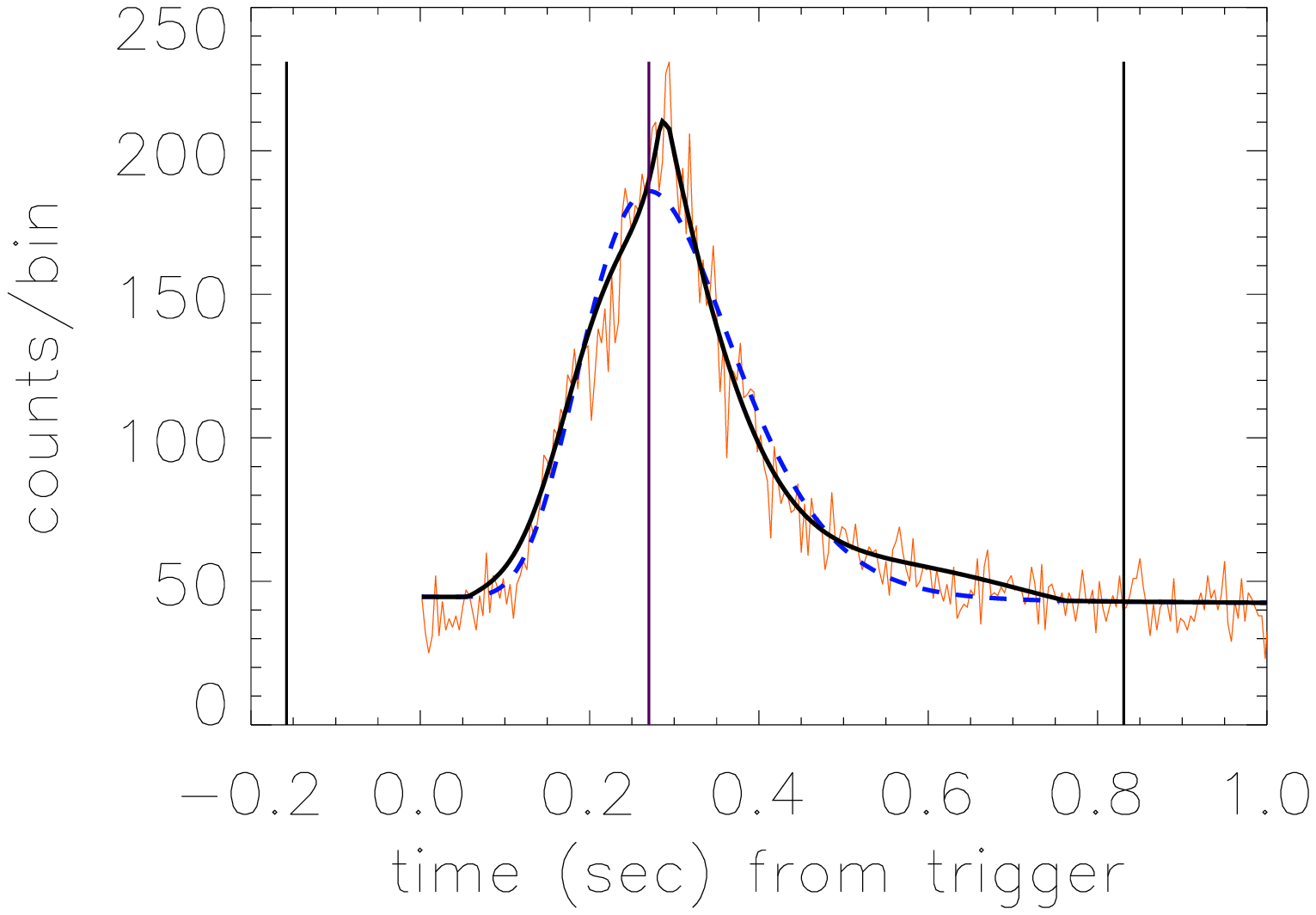}
    \vspace*{-0.5 cm}
  \end{subfigure}
  \begin{subfigure}[b]{0.4\textwidth}
    \includegraphics[width=\textwidth]{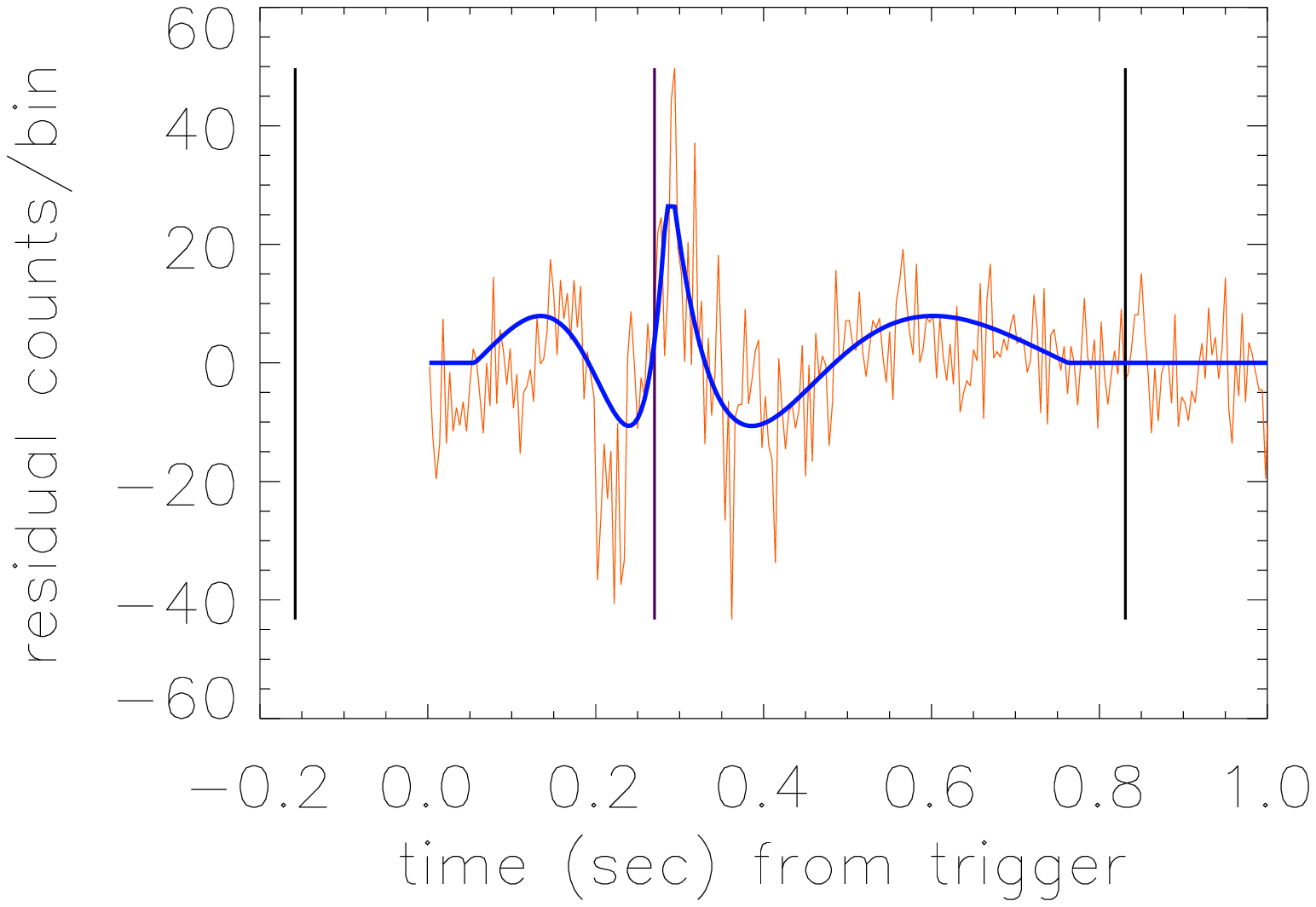}
    \vspace*{-0.5 cm}
  \end{subfigure}
  \end{center}
 \vspace*{-0.1 cm}
  \caption{Blended SGRB pulse BATSE 2896. Pulse fit (left) and residual fit (right).}
  \label{fig2}

\vspace*{-0.4 cm}
\begin{center}
  \begin{subfigure}[b]{0.4\textwidth}
    \includegraphics[width=\textwidth]{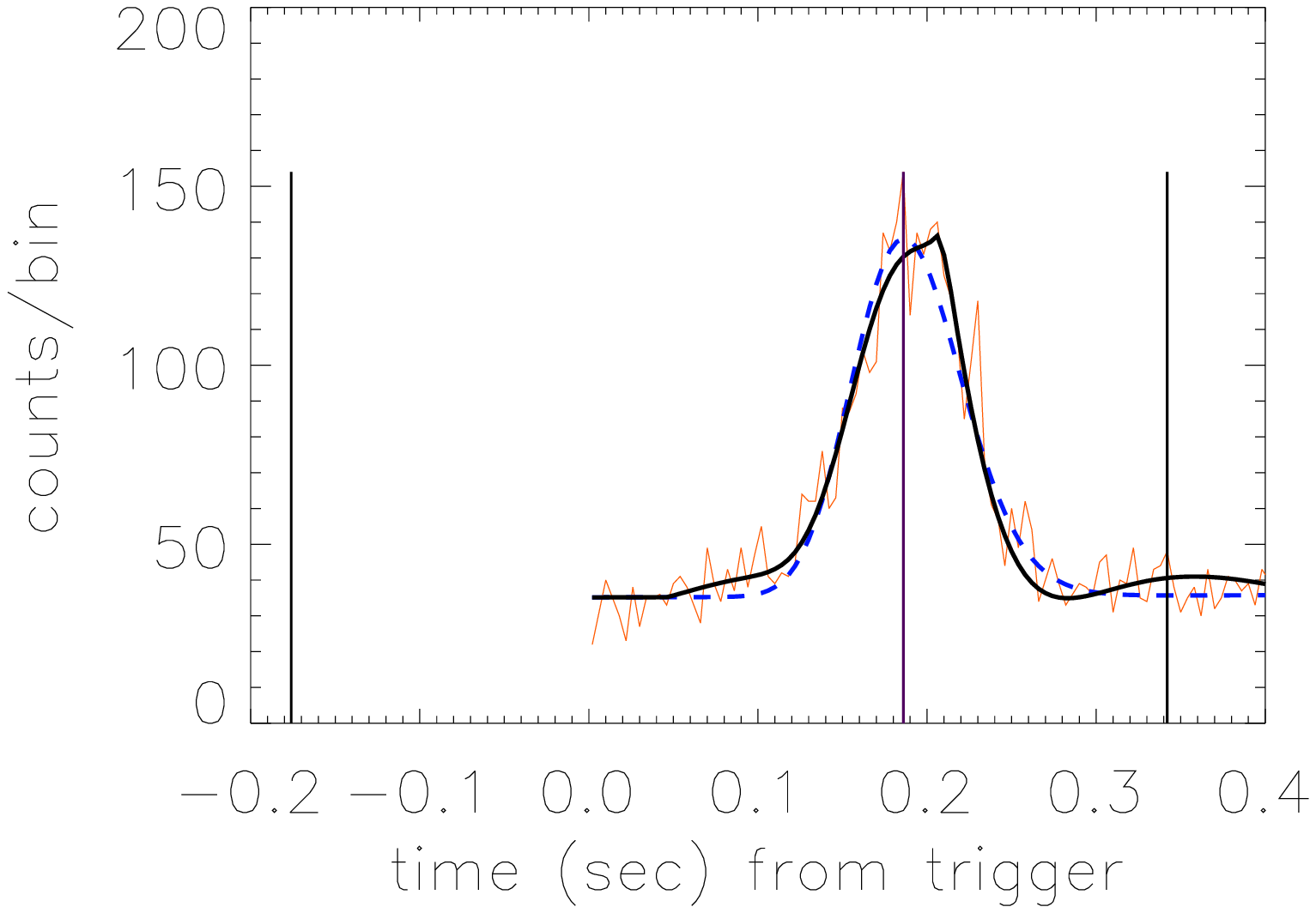}
    \vspace*{-0.5 cm}
  \end{subfigure}
  \begin{subfigure}[b]{0.4\textwidth}
    \includegraphics[width=\textwidth]{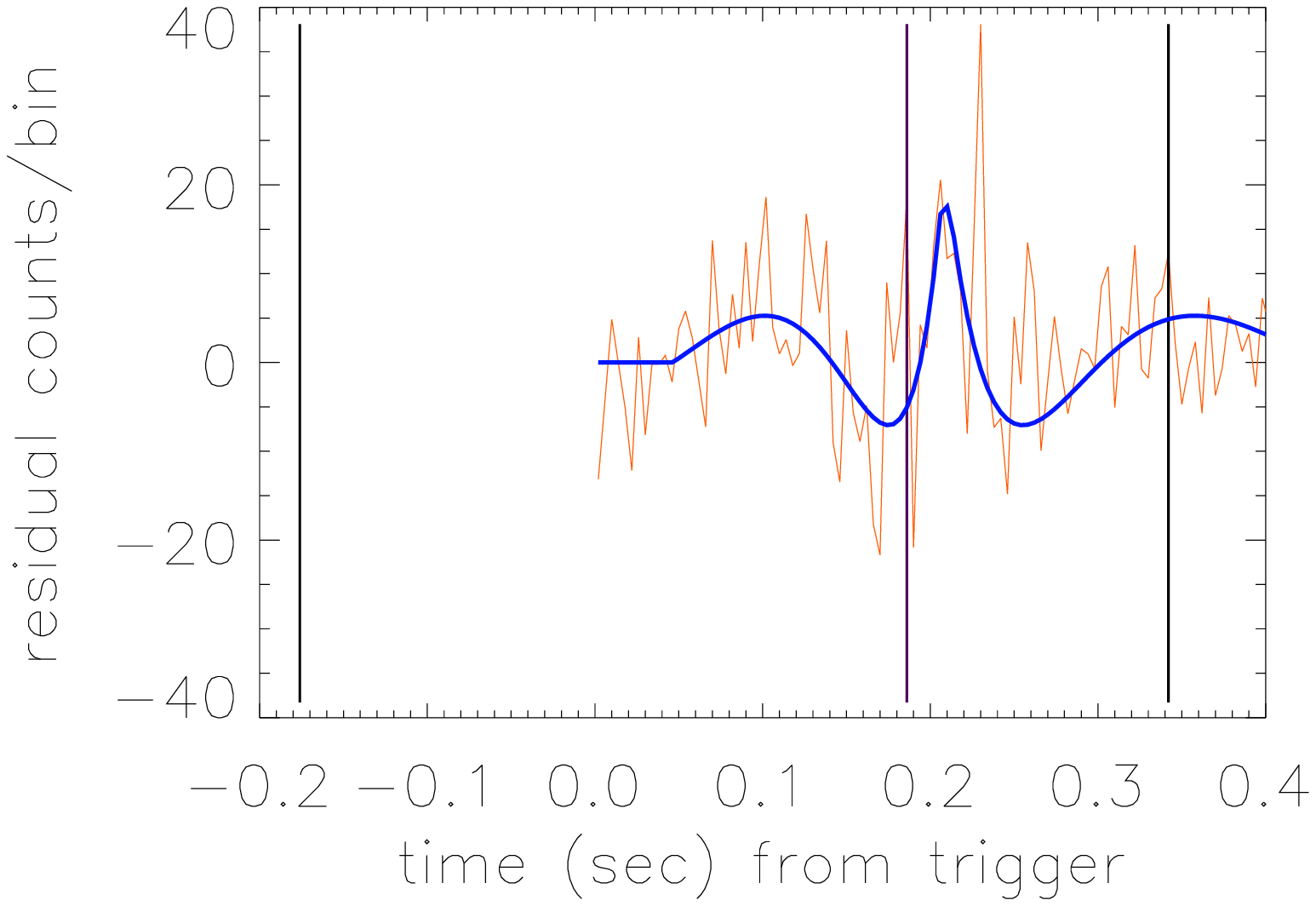}
    \vspace*{-0.5 cm}
  \end{subfigure}
  \end{center}
 \vspace*{-0.1 cm}
  \caption{Structured SGRB pulse BATSE 5564. Pulse fit (left) and residual fit (right).}
  \label{fig3}

\vspace*{-0.4 cm}
\begin{center}
  \begin{subfigure}[b]{0.4\textwidth}
    \includegraphics[width=\textwidth]{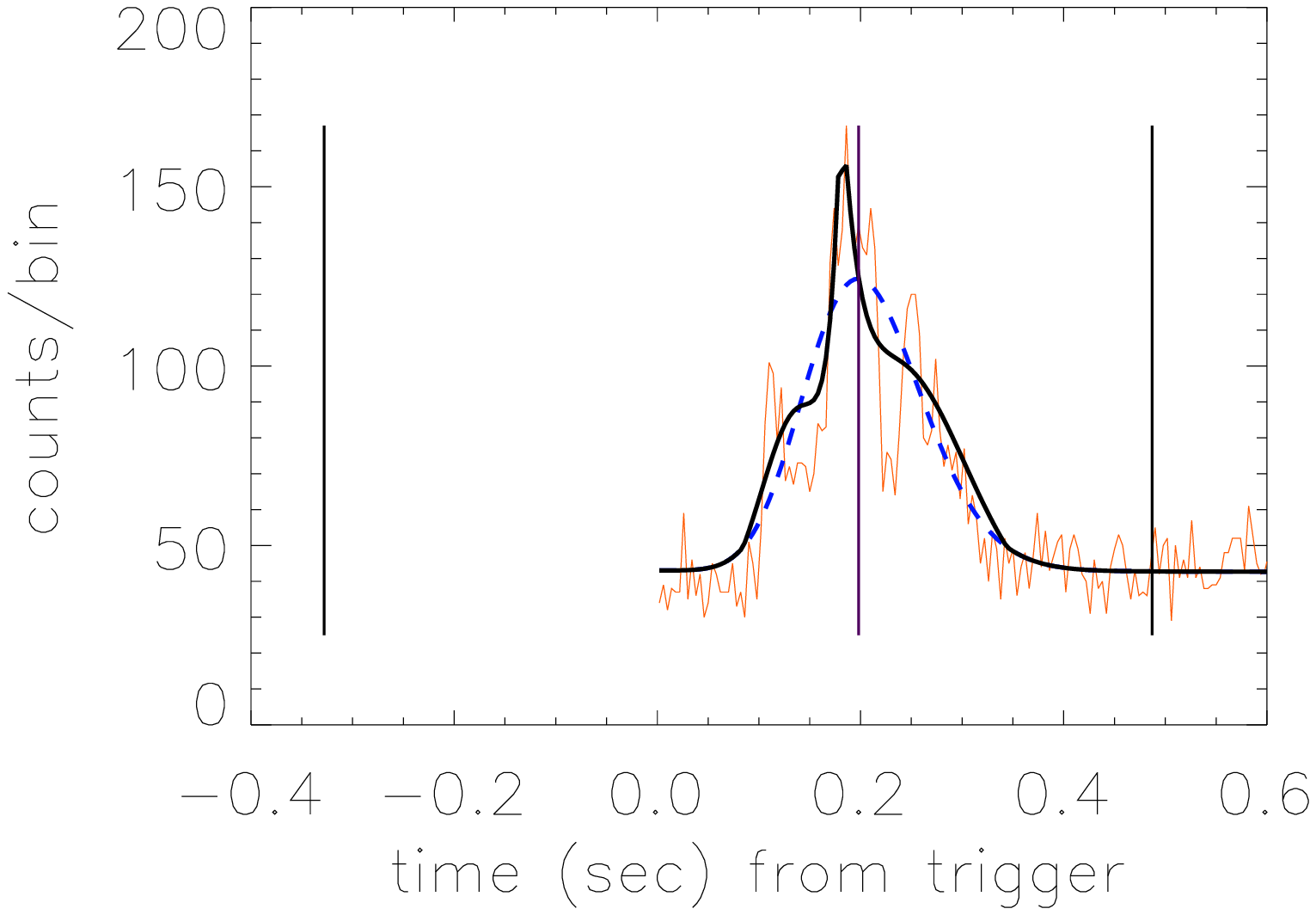}
    \vspace*{-0.5 cm}
  \end{subfigure}
  \begin{subfigure}[b]{0.4\textwidth}
    \includegraphics[width=\textwidth]{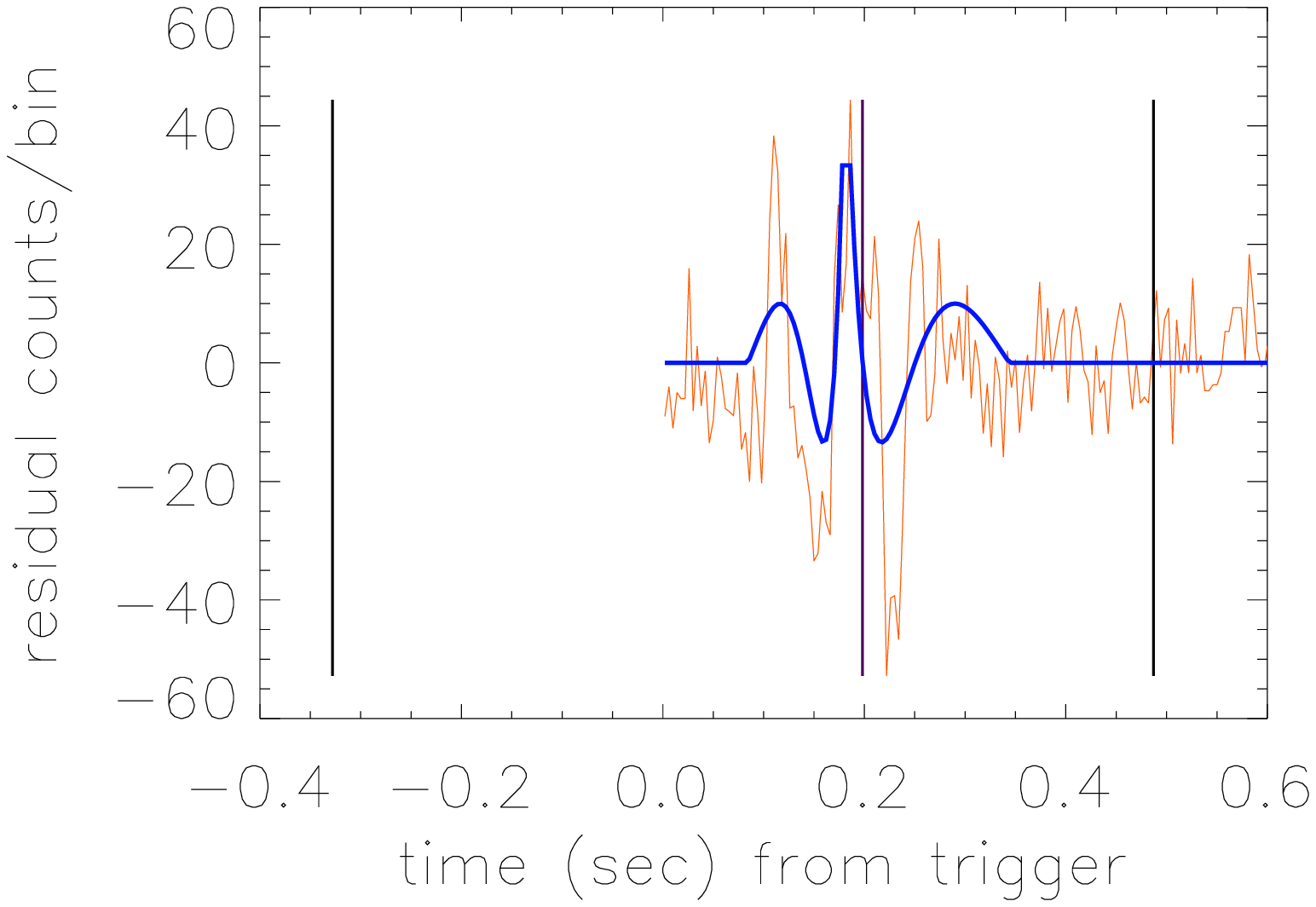}
    \vspace*{-0.5 cm}
  \end{subfigure}
  \end{center}
 \vspace*{-0.1 cm}
  \caption{Complex SGRB pulse BATSE 4955. Pulse fit (left) and residual fit (right).}
  \label{fig4}
\end{figure}

Examples of SGRB pulses containing different amounts of structure are
shown for BATSE SGRBs 0373 (simple), 2896 (blended),
5564 (structured), and 4955 (complex) in Figs. \ref{fig1} through \ref{fig4}. In each figure,
the left panel demonstrates the fit obtained with the \cite[Norris et al. (2005)]{Norris05} 
pulse shape (dashed blue line) and the \cite[Norris et al. (2005)]{Norris05} pulse shape 
combined with the \cite[Hakkila \& Preece (2014)]{Hakkila14} residual structure (solid black line).  
The right panel indicates the residuals once the \cite[Norris et al. (2005)]{Norris05} fit
has been removed, overlaid by the \cite[Hakkila \& Preece (2014)]{Hakkila14} residual fit (solid blue line).


Detector signal-to-noise ratio (S/N) and temporal resolution are capable of smearing out intrinsically 
complex GRB structures and of causing GRB pulses to appear as either monotonic shapes 
augmented by the triple-peaked structure or as simple monotonic shapes. Thus a 
GRB pulse's appearance is a combination of intrinsic structures and instrumental 
smearing effects. The effects of S/N on both SGRB and LGRB pulse classification have been demonstrated 
by \cite[Hakkila et al. (2018a)]{Hakkila18b} and \cite[Hakkila et al. (2018b)]{Hakkila18b}.

\section{Constraints Imposed by Time-Reversed and Stretched Residuals}

Since instrumental effects can make it hard to delineate GRB pulse structure from noise,
interpretation of GRB pulse physics is best understood through the study of bright GRB pulses.
\cite[Hakkila et al. (2018b)]{Hakkila18b} studied six of the brightest BATSE LGRB pulses
and demonstrated that the residual structure model employed previously was too simple
and incomplete for describing the residuals of these pulses because the residual structure extends
far beyond the temporal boundaries containing the three peaks.
Furthermore, the extended wavelike structure is shown to have strange characteristics: 
it is both {\em time-reversible} and {\em stretched}
around a {\em time of reflection}. In other words, the pulse residuals following 
the time of reflection have a memory of the residuals preceding it, but these events are
repeated in reverse order after undergoing a dilation at the time of reflection.

\begin{figure}[H]
\begin{center}
  \begin{subfigure}[b]{0.4\textwidth}
    \includegraphics[width=\textwidth]{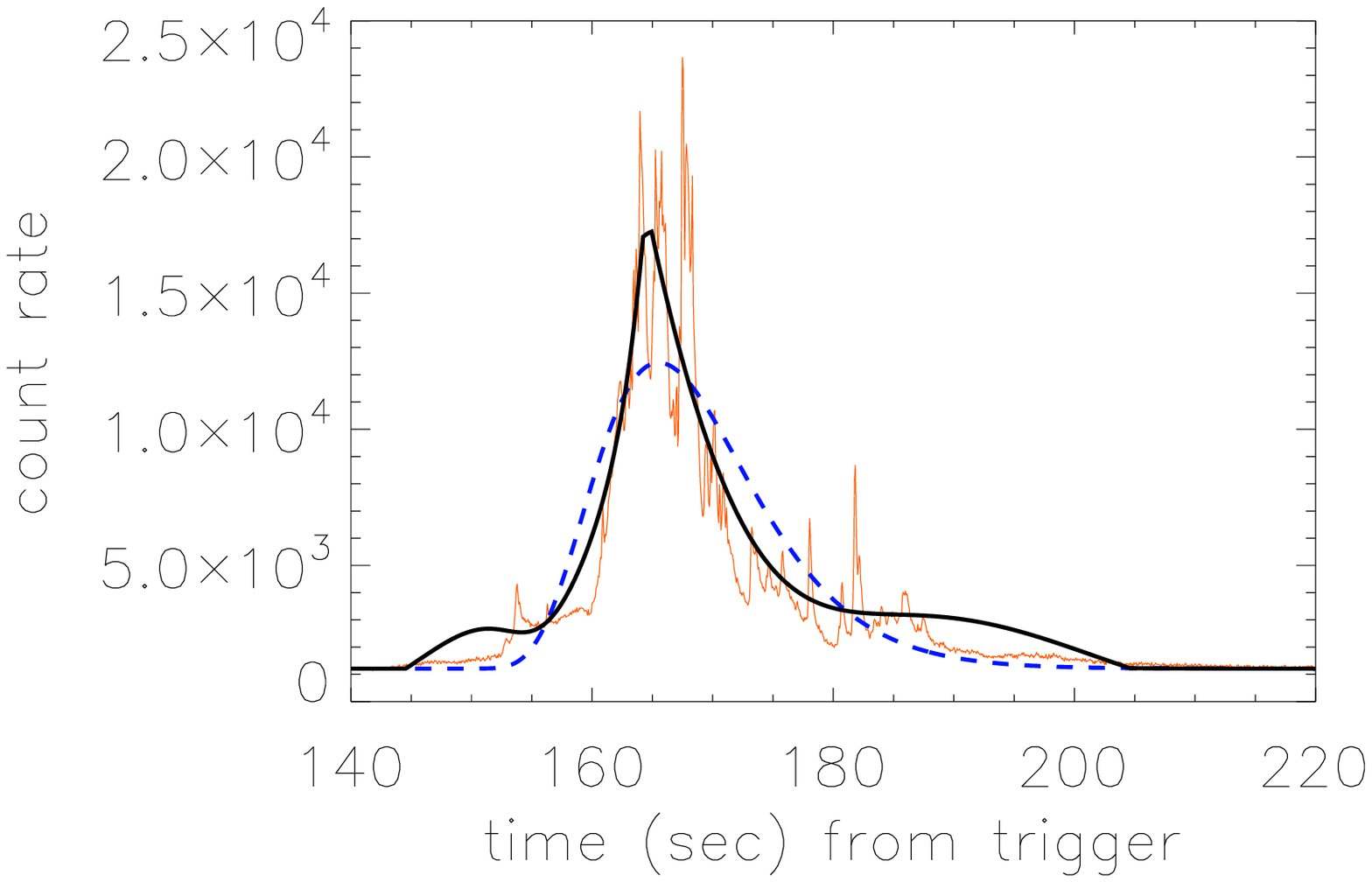}
  \end{subfigure}
  \begin{subfigure}[b]{0.4\textwidth}
    \includegraphics[width=\textwidth]{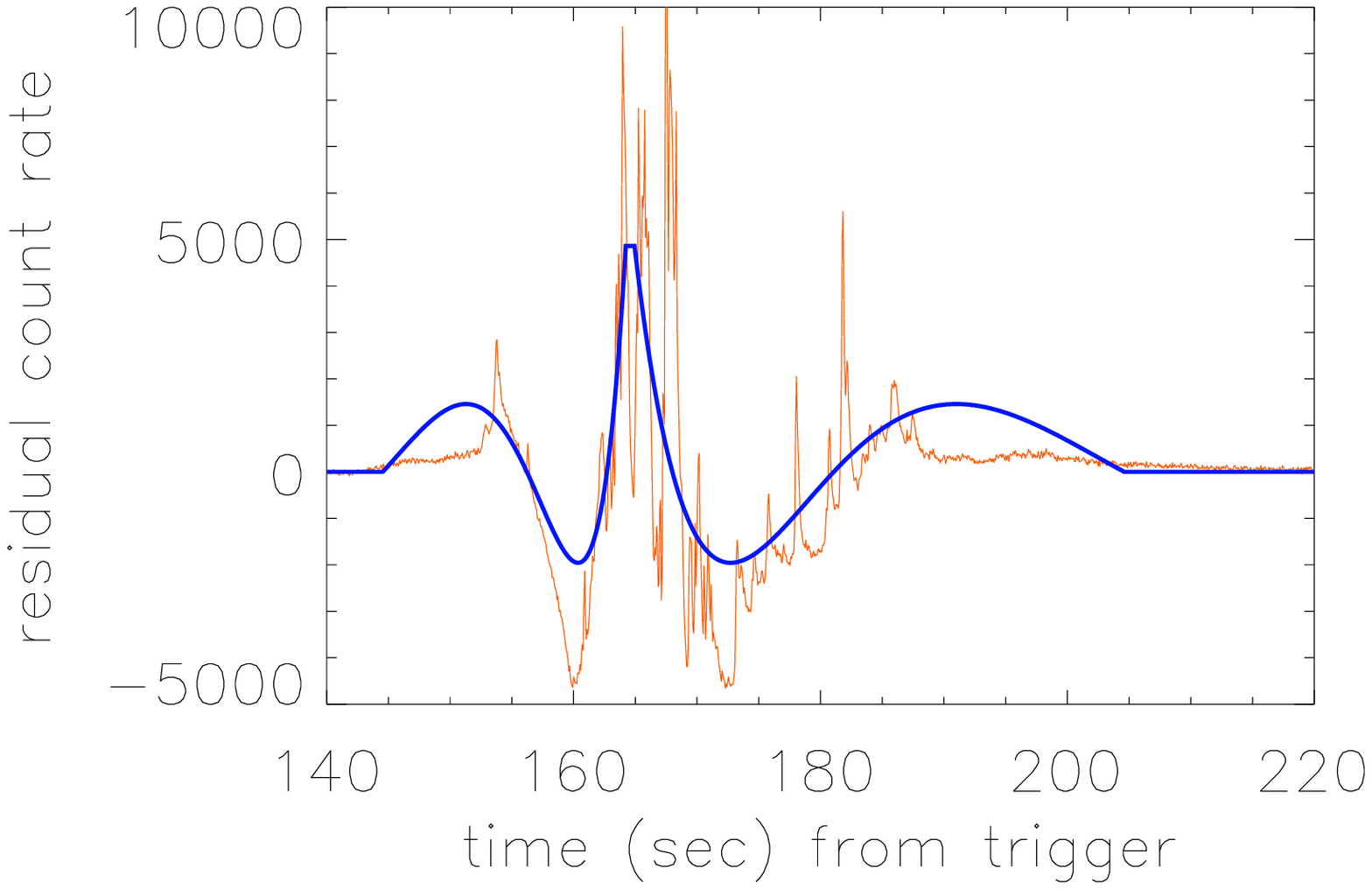}
  \end{subfigure}
  \end{center}
  \caption{LGRB BATSE 7301p2, a complex, extremely bright pulse. Pulse fit (left) and residual fit (right).}
  \label{fig5}

\begin{center}
  \begin{subfigure}[b]{0.4\textwidth}
    \includegraphics[width=\textwidth]{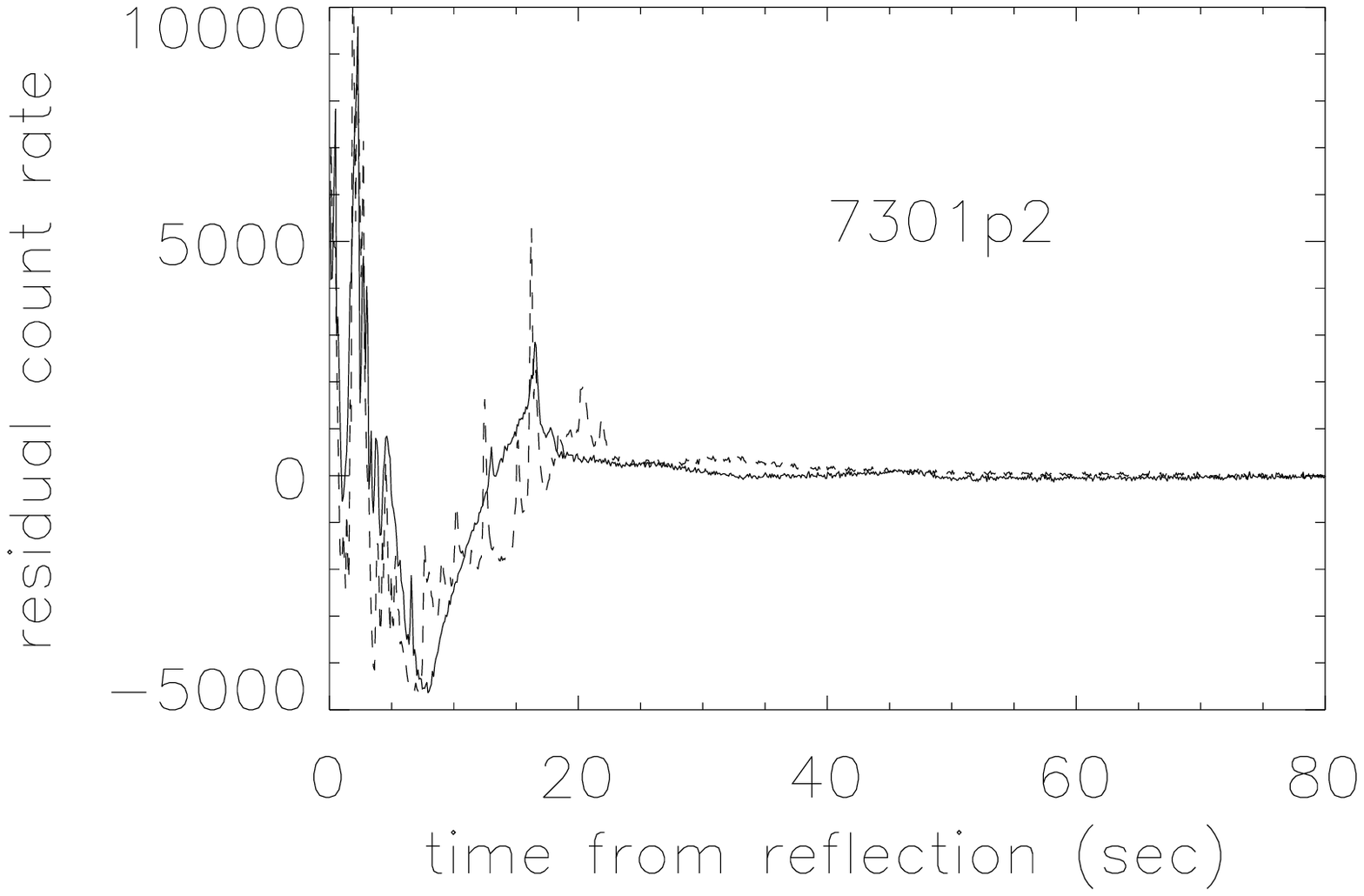}
  \end{subfigure}
  \begin{subfigure}[b]{0.4\textwidth}
    \includegraphics[width=\textwidth]{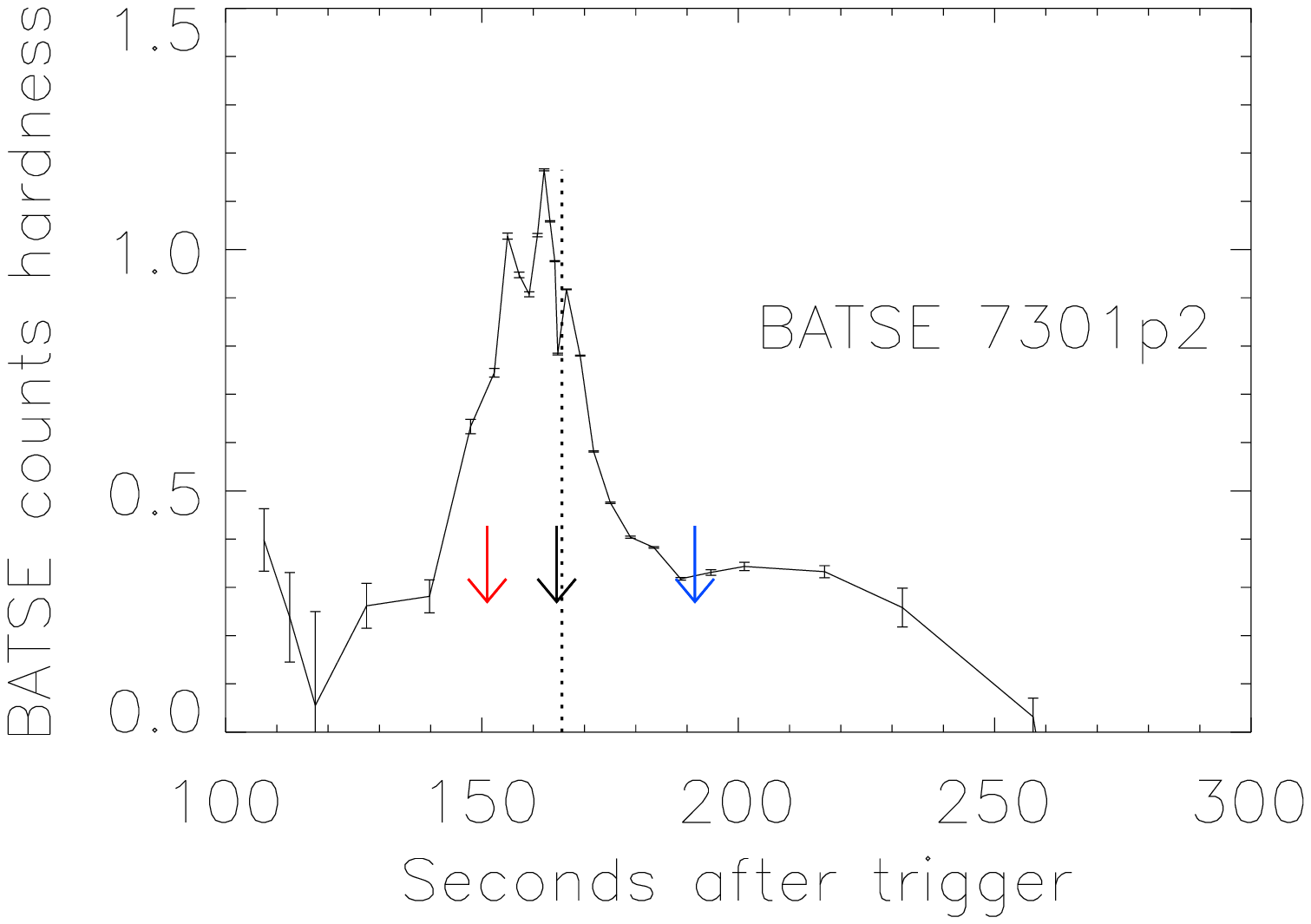}
  \end{subfigure}
  \end{center}
  \caption{LGRB BATSE 7301p2. Time-reversed and stretched residuals (left) and spectral hardness evolution (right).}
  \label{fig6}
\end{figure}

An example of this is shown for LGRB BATSE pulse 7301 p2 in Fig. \ref{fig5} and Fig. \ref{fig6}.
The left panel of this figure shows the model fits for the pulse, both including (solid black line)
and excluding (dashed blue line). The residual model is able to fit part of the residual light curve,
but is inadequate in identifying and fitting all of the structure (right panel of Fig. \ref{fig5}).
The left panel of Fig. \ref{fig6} uses a new approach that recognizes the time-reversed and
stretched structure of the residual model without being dependent on that model's functional form.
The residuals are folded over in time and stretched until they line up with one another,
so the residuals prior to the time of reflection (solid line) are shown overlaid by the 
time-reversed and stretched residuals preceding the time of reflection (dashed line).
Further evidence that these residuals are linked together in a chain, rather than
distributed randomly, is shown in the hardness evolution plot of the pulse (Fig. \ref{fig6}).
Here the pulse hardness generally evolves from hard-to-soft, with a re-hardening
at each residual peak.

Time-reversed and stretched pulse residuals place remarkably strict constraints
on GRB models. They couple events that happen at the beginning of a pulse
with those that happen towards the end, but they further indicate that the conditions
responsible for creating the pulse structure must repeat in reverse order and be time-dilated.
We demonstrate two simple models in which pulse light curves with the observed 
characteristics might be created by jetted GRB material. We note that these models
are driven solely by kinematics and geometry rather than by a specific radiation 
mechanism. 

\begin{figure}[H]
\vspace*{-0.8 cm}
\begin{center}
  \begin{subfigure}[b]{0.4\textwidth}
    \includegraphics[width=\textwidth]{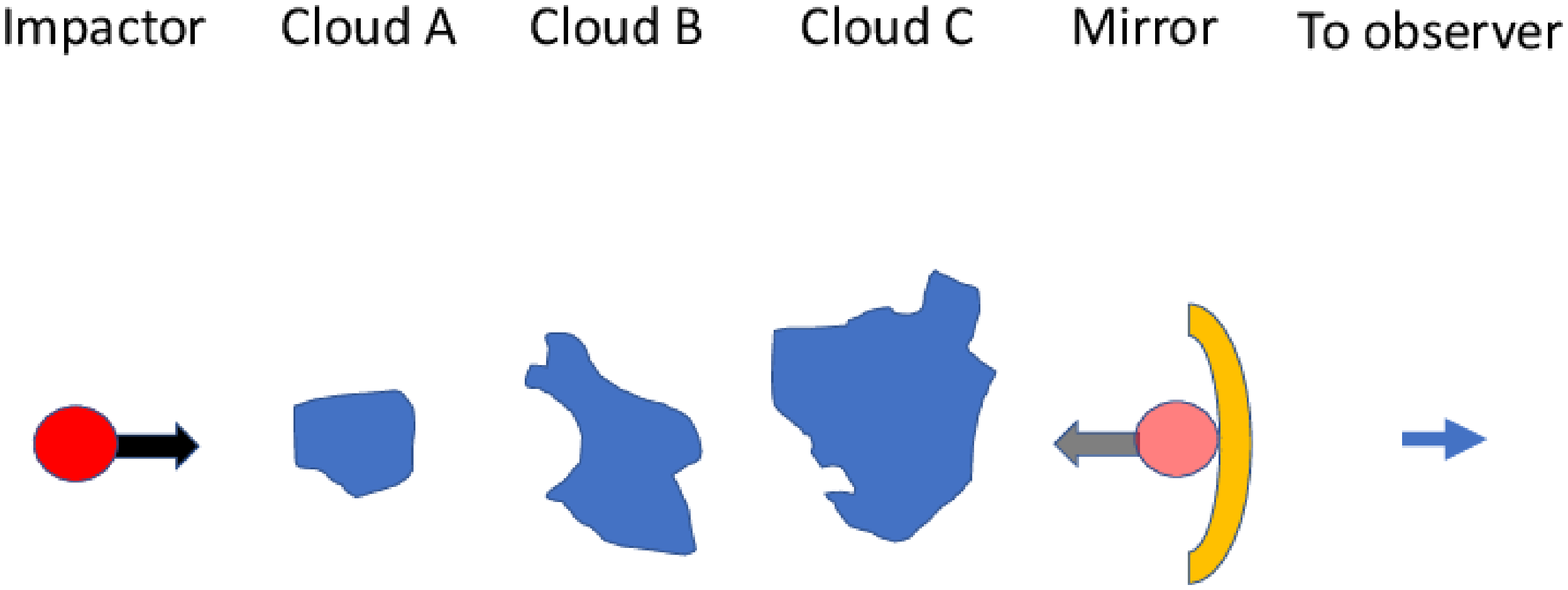}
  \end{subfigure}
  \begin{subfigure}[b]{0.4\textwidth}
    \includegraphics[width=\textwidth]{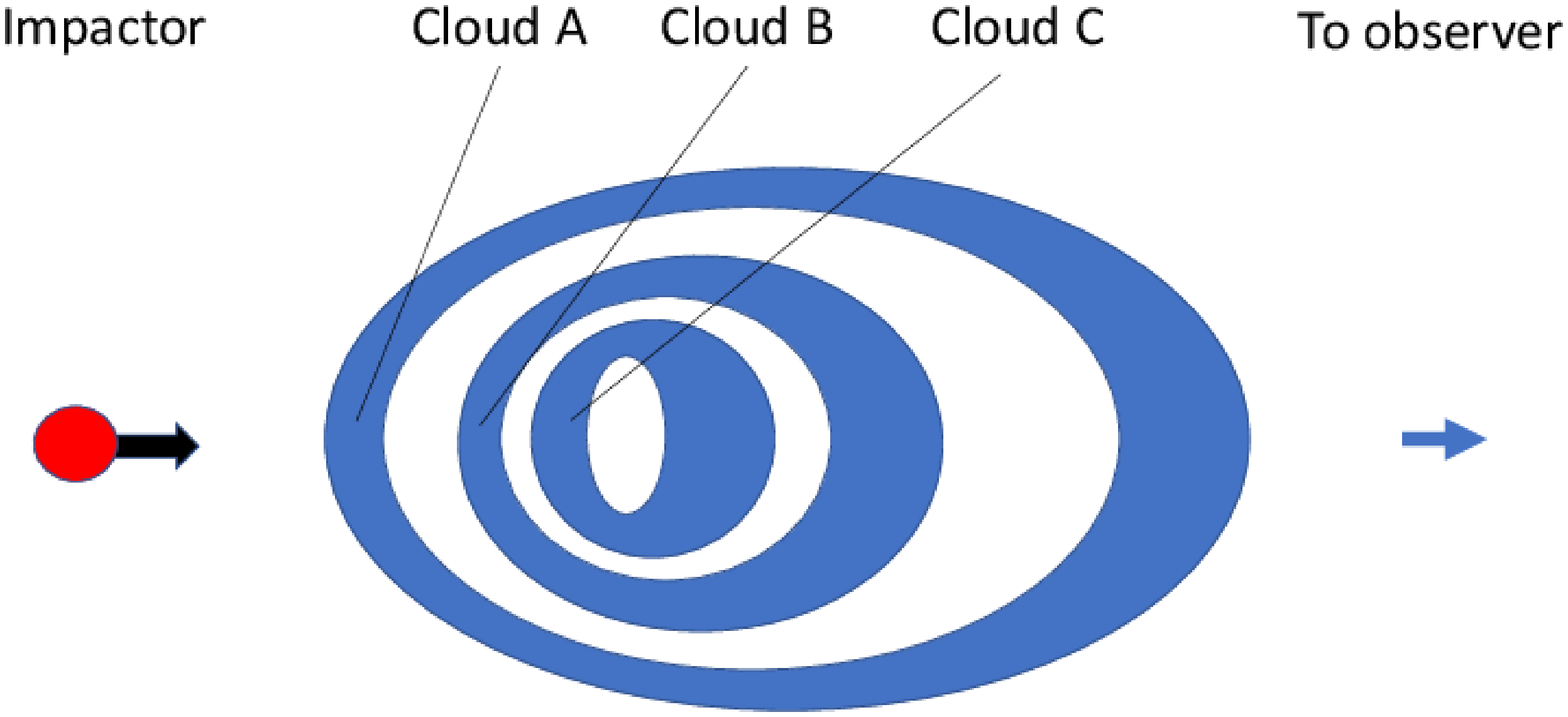}
  \end{subfigure}
  \end{center}
  \caption{Two possible kinematic models for explaining the time-reversed and stretched residuals found in GRB pulse light curves: the mirror model (left panel) and bilaterally-symmetric model (right panel).}
  \label{fig7}
\end{figure}

The first {\em mirror model}, shown in the left panel of Fig. \ref{fig7} consists of a
relativistically-moving {\em impactor} (shown as a red circle) ejected from the central engine.
This impactor might be a soliton (shock wave) or a plasma blob that interacts
with other plasma clouds in the jet (in blue) to produce radiation.
The impactor slows upon striking a mirror (in yellow; presumably the jet head),
which allows the clouds to catch up with it in the opposite order. The blueshifted initial
motion of the impactor through the clouds produces beamed emission (A.B.C) followed by 
emission that is less-strongly blueshifted (C...B...A).

The second {\em bilaterally-symmetric model}, shown in the right panel of Fig. \ref{fig7},
is composed of clouds distributed in a bilaterally-symmetric fashion along the impactor's path, 
producing the beamed (A.B.C.C...B...A) emission pattern.

\section{Constraints Imposed by the Rarity of Pulses}

The standard definition of a pulse refers to a single-peaked monotonic bump. 
Using this definition, observers are misled into thinking that a typical GRB 
generally contains many pulses.
The recognition that peaks are linked temporally (such that the time-reversed and
stretched residuals can be used to identify all the peaks associated with a single pulse),
allows the number of pulses in a GRB to be reduced dramatically. 

The recent study of \cite[Hakkila et al. (2018b)]{Hakkila18b} finds that
$90\%$ of SGRBs are single-pulsed, and most of the remaining $10\%$ are double-pulsed.
Thus the mechanism producing SGRBs generally does so in the form of
a single structured pulse, but this can also occur less frequently as two or maybe three pulses.
Since SGRBs are produced by colliding neutron stars, it seems unlikely that each
interaction is capable of producing more than a single blast wave. We thus
turn to GRB geometry to explain multi-pulsed bursts.

\section{An SGRB Model that Accounts for Pulse Structure and Rarity}

We can use the constraints imposed by GRB pulse structure and by the rarity of
multi-pulsed SGRBs, in conjunction with theoretical models of merging neutron
stars, to improve physical models. Standard models of merging neutron stars 
suggest that they produce a thick accretion disk with a tail extending behind it as 
it rotates ({\em e.g.} \cite[Rosswog, Piran, \& Nakar (2013)]{Rosswog13}). 
The timescale for 
the existence of this disk is very short ($< 20$ ms). From the perspective of our model,
we can consider the radial distribution of the disk to be the `jet' and
density variations in the disk itself to comprise the distribution of clouds within the jet.

\begin{figure}[H]
\begin{center}
  \begin{subfigure}[b]{0.4\textwidth}
    \includegraphics[width=\textwidth]{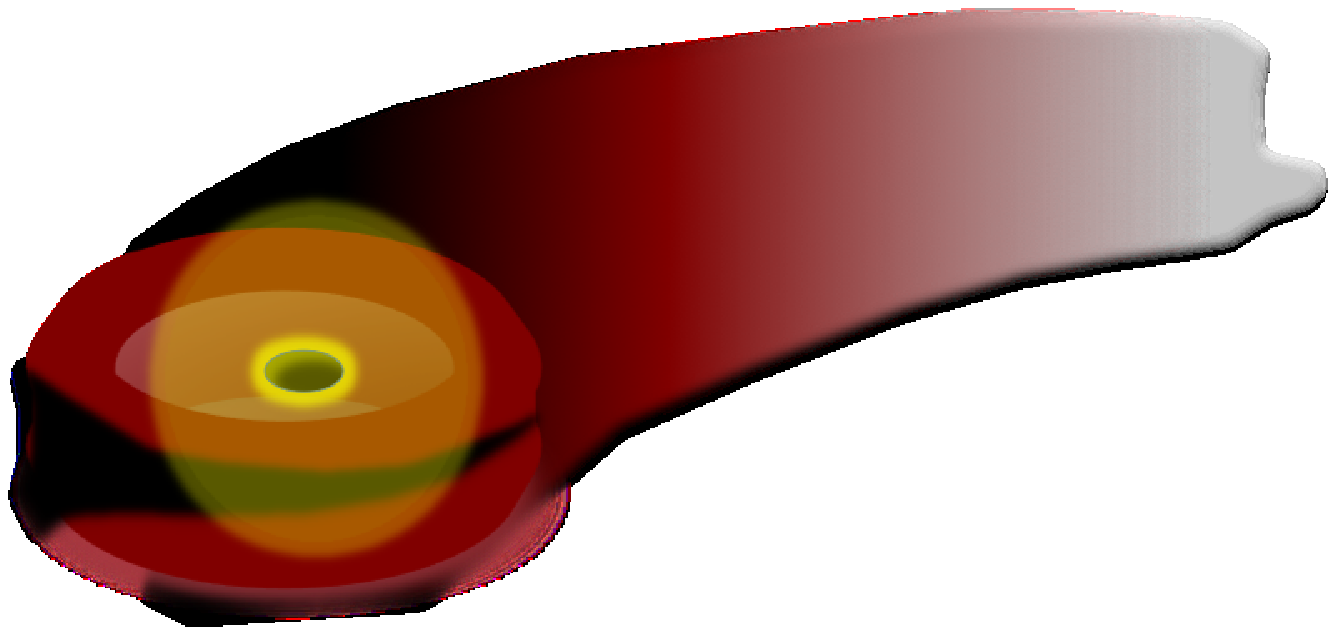}
  \end{subfigure}
  \begin{subfigure}[b]{0.4\textwidth}
    \includegraphics[width=\textwidth]{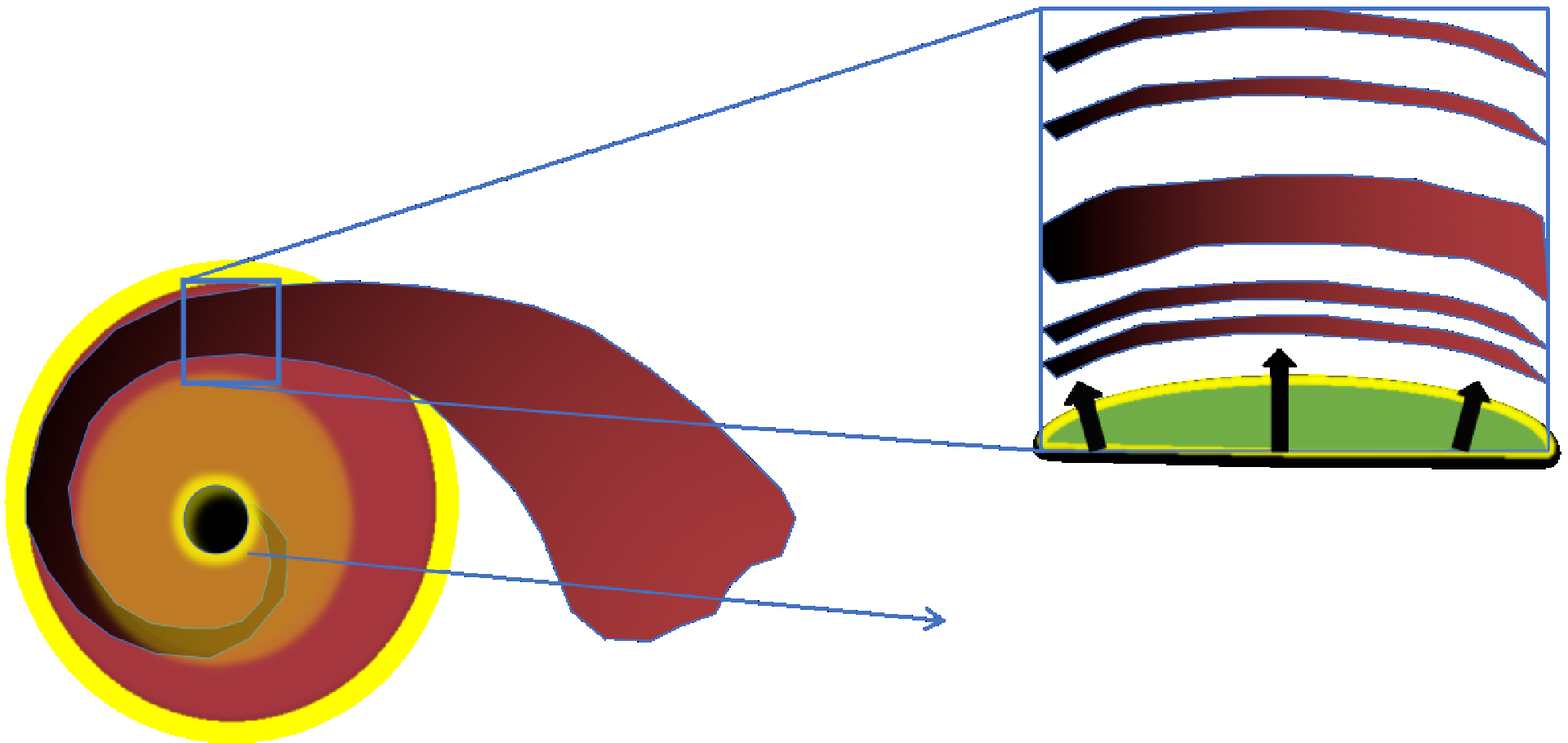}
  \end{subfigure}
  \end{center}
  \caption{3D model of ns-ns merger as seen from the side (left) and top (right) views. Most lines-of-sight produce single-pulsed SGRBs, but a line-of-sight through the accretion tail will produce two pulses. The accretion disk structure is similar to that found in the axially-symmetric model shown in Figure \ref{fig7}.}
  \label{fig8}
\end{figure}

In order to match our observations, the merging neutron stars likely
produces a soliton at the moment of black hole formation, and this impactor
expands spherically outward (denoted by the yellow sphere in Fig. \ref{fig8}).
This model can reproduce the time-reversed
and stretched residuals found in SGRB pulse light curves if the radial distribution 
of material is bilaterally-symmetric (seen in the enlargement of the accretion disk
radial structure found on the far right side of Fig. \ref{fig8}). We note that double-pulsed
bursts can occur if the accretion tail is pointed along the observer's line-of-sight,
so that the observer sees two emitted pulses each with similar time-reversed
and stretched structures. The timescale of each pulse is essentially the
light travel time of the disk and of the tail, and the interpulse separation is 
essentially the light travel time of the gap between the disk and the tail.

This attempt to model SGRBs is among the first to 
incorporate constraints imposed by observations of GRB pulses.
Other models are also possible, but each must be consistent
with the pulse observations. We continue to study and explore these models
for both SGRBs and LGRBs.

\end{document}